\documentstyle[12pt]{article}
\begin{document}
\hfill{NCKU-HEP-97-05}
\vskip 0.5cm
\begin{center}
{\large {\bf A modified BFKL equation with $Q$ dependence}}
\vskip 1.0cm
Jyh-Liong Lim
\vskip 0.3cm
Department of Electrophysics, National Chiao-Tung University, \par
Hsin-Chu, Taiwan, Republic of China
\vskip 0.3cm
Hsiang-nan Li
\vskip 0.3cm
Department of Physics, National Cheng-Kung University, \par
Tainan, Taiwan, Republic of China
\end{center}
\vskip 1.0cm

\vskip 1.0cm
\centerline{\bf Abstract}
\vskip 0.3cm
We propose a modified Balitsky-Fadin-Kuraev-Lipatov (BFKL) equation for
the summation of large $\ln(1/x)$,
$x$ being the Bjorken variable, which contains an extra dependence on
momentum transfer $Q$ compared to the conventional BFKL equation. This
$Q$ dependence comes from the phase space constraint for soft real gluon
emissions in the derivation of the BFKL kernel. The modified equation
gives a slower rise of the gluon distribution function with $1/x$
for smaller $Q$, such that the predictions for the structure functions
$F_{2,L}(x,Q^2)$ involved in deep inelastic scattering 
match the HERA data. These results are similar to those obtained from the
Ciafaloni-Catani-Fiorani-Marchesini equation, which embodies both the
$\ln Q$ and $\ln(1/x)$ summations. The reason that the
conventional BFKL equation can not be applied to the low $Q$ (soft pomeron
dominant) region is supplied.

\newpage
\centerline{\large\bf I. INTRODUCTION}
\vskip 0.5cm

The deep inelastic scattering (DIS) experiments performed at HERA 
\cite{H1} have stimulated many intensive discussions on the behavior of
the structure function $F_2(x,Q^2)$ at small Bjorken variable $x$ for
various momentum transfer $Q$. One of the important subjects is to explain
the data of $F_2$ using the known evolution equations. In the low $x$
region with $\ln(1/x) \gg \ln Q$, the appropriate tool is the
Balitsky-Fadin-Kuraev-Lipatov (BFKL) equation \cite{BFKL} which sums large
$\ln(1/x)$ in the gluon distribution function to all orders. Since the
BFKL equation does not depend on momentum transfer explicitly, its
predictions for the gluon distribution function and for $F_2$ exhibit a
weak $Q$ dependence. However, the HERA data show that the rise of $F_2$ at
small $x$ varies with $Q$ in a more sensitive way: The ascent is rapid for
large $Q$, which is attributed to hard (or BFKL) pomeron
contributions. The slow ascent at small $Q$, which can not be explained by
the BFKL equation, is interpreted as the consequence of soft pomeron
exchanges.

To understand the $Q$-dependent rise of the gluon distribution function and
the structure function, the
Dokshitzer-Gribov-Lipatov-Altarelli-Parisi (DGLAP) equation \cite{GLAP} for
the $\ln Q$ summation should be included somehow. One may resort to the
DGLAP equation directly, because the relevant splitting function
$P_{gg}\propto 1/x$ also gives an increase at small $x$. However, the DGLAP
rise is milder, and a steep nonperturbative input of the gluon distribution
function at low $Q$ must be adopted. An alternative is the
$p_T$-factorization theorem \cite{J}, which has been shown to be equivalent
to the collinear (mass) factorization, the framework on which the DGLAP
equation is based. By including the next-to-leading $\ln(1/x)$ from singlet
quark contributions, the rise becomes faster, and a flat input is proposed
\cite{EHW}. The double scaling tendency exhibited by the predictions from
the DGLAP equation with the next-to-leading logarithms taken into
account and by the data of $F_2$ has been explored \cite{BF}. The
conclusion also favors a flat nonperturbative input. The above
investigations imply that the structure function at low $Q$ is dominated
by soft pomeron contributions, and the large-$Q$ behavior is the
result of the DGLAP evolution.

Another possibility is the Ciafaloni-Catani-Fiorani-Marchesini (CCFM)
equation \cite{CCFM}, which embodies the BFKL equation for small $x$ (and
intermediate $Q$) and the DGLAP equation for large $Q$ (and intermediate
$x$). To derive the CCFM equation, the angular ordering of rung gluons in
a ladder diagram is assumed, which can be regared as a mixture of the
rapidity ordering for the BFKL equation and the transverse momentum
ordering for the DGLAP equation. The CCFM equation has been studied in
details in \cite{KMS1}. It is found that the resultant gluon distribution
function shows the desired $Q$ dependence, and the predictions for
$F_2$ are in agreement with the HERA data. This equation, though
considering only the gluon contributions, predicts a steeper rise than
the DGLAP equation does, since the running coupling constant
involved in the splitting function is evaluated at the gluon transverse
momentum, instead of at $Q$.

The above analyses, though consistent with the data, did not address the
issue why the BFKL equation can not be applied to DIS in the low $Q$ region,
where it is supposed to be more appropriate because of $\ln(1/x) \gg \ln Q$.
In this paper we shall derive the BFKL equation in a more careful way from
the viewpoint of the Collins-Soper-Sterman resummation technique \cite{CS}.
By truncating the loop momenta of real gluons at $Q$ in the evaluation of
the evolution kernel, a modified equation is obtained. Since real gluon
emissions are responsible for the BFKL rise, the rise is rendered slower at
smaller $Q$, where the phase space for real gluon emissions is more
restricted. We argue that this is the reason the conventional
BFKL equation, overestimating real gluon contributions, fails 
in the low $Q$ region. The modified equation is simpler than the
CCFM equation, and explains the HERA data of the structure functions
$F_{2,L}$ well.

In Sect. II we derive the $Q$-dependent BFKL equation using the
resummation technique, and show that it approaches the conventional one as
$Q\to \infty$. In Sect. III the evolution of the gluon distribution function
in $x$ predicted by the modified BFKL equation is compared with those by
the DGLAP equation, the conventional BFKL equation, and the CCFM equation.
The modified BFKL equation is solved numerically, and the results of the
gluon distribution function and of the structure functions $F_{2,L}$ are
presented in Sect. IV. Section V is the conclusioin, and the Appendix
contains the details of the involved calculations.

\vskip 1.0cm
\centerline{\large \bf II. FORMALISM}
\vskip 0.5cm

In \cite{L2,L3} we have proposed a unified derivation of the known
evolution equations that sums various large logarithms based on the
resummation technique \cite{CS}, such as the DGLAP,
BFKL, CCFM, and Gribov-Levin-Ryskin (GLR) equation \cite{GLR}. The GLR
equation includes the annihilation effect of two gluons into one gluon in
the region with $Q$ and $1/x$ being simultaneously large. Briefly speaking,
the resummation technique relates the derivative of a parton distribution
function to a new function involving a new vertex \cite{CS,L1}. The new
function is then expressed as a factorization formula involving the
subdiagram containing the new vertex and the original distribution function.
Evaluating the subdiagram according to the specific orderings of radiative
gluons simply, the derivative of the distribution function becomes the
corresponding evolution equation. The complexity of the conventional
derivation of the evolution equations is thus greatly reduced.

In this section we apply the above formalism to DIS of a proton with
light-like momentum $P^\mu=P^+\delta^{\mu +}$ in the small $x$ limit, where
$x=-q^2/(2P\cdot q)=Q^2/(2P\cdot q)$ is the Bjorken variable, $q$ being
the momentum of the photon.
The unintegrated gluon distribution function $F(x,k_T)$,
describing the probability of a gluon carrying a longitudinal momentum
fraction $x$ and transverse momentum $p_T$, is defined by
\begin{eqnarray}
F(x,p_T)&=&\frac{1}{P^+}\int\frac{dy^-}{2\pi}\int\frac{d^2y_T}{4\pi}
e^{-i(xP^+y^--{\bf p}_T\cdot {\bf y}_T)}
\nonumber \\
& &\times\langle P| F^+_\mu(y^-,y_T)F^{\mu+}(0)|P\rangle\;,
\label{deg}
\end{eqnarray}
in the axial gauge $n\cdot A=A^+=0$, $n=\delta^{\mu-}$ being a vector on
the light cone. The ket $|P\rangle$ denotes the incoming proton, and
$F^+_\mu$ is the field tensor. Averages over the spin and color are
understood. To implement the resummation technique, we allow $n$ to
vary away from the light cone ($n^2\not=0$) first. It will
be shown that the BFKL kernel turns out to be $n$-independent. After
deriving the evolution equation, $n$ is brought back to the light cone,
and the definition of the gluon distribution function coincides with
the standard one. That is, the arbitrary vector $n$ appears only at the
intermediate stage of the derivation and as an auxiliary tool of the
resummation tecchnique.

Though $F$ at $x\to 0$ does not depend on
$P^+$ explicitly as shown later, it varies with $P^+$ through the momentum 
fraction implicitly, which is proportional to $(P^+)^{-1}$. Hence,
the derivative $P^+dF/dP^+$ can be written as
\begin{equation}
P^+\frac{d}{dP^+}F(x,P_T)\equiv -x\frac{d}{dx}F(x,P_T)\;.
\end{equation}
Because of the scale invariance of $F$ in $n$, as indicated by
the gluon propagator, $(-i/l^2)N^{\mu\nu}(l)$, with
\begin{equation}
N^{\mu\nu}=g^{\mu\nu}-\frac{n^\mu l^\nu+n^\nu l^\mu}
{n\cdot l}+n^2\frac{l^\mu l^\nu}{(n\cdot l)^2}\;,
\label{gp}
\end{equation}
$F$ must depend on $P$ via the ratio $(P\cdot n)^2/n^2$. 
We have the chain rule relating $P^+d/dP^+$ to $d/dn$ \cite{CS,L1},
\begin{eqnarray}
P^+\frac{d}{dP^+}F=-\frac{n^2}{v\cdot n}v_\beta\frac{d}{dn_\beta}F\;,
\label{cph}
\end{eqnarray}
$v_\beta=\delta_{\beta +}$ being a vector along $P$.
The operator $d/dn_\beta$ appplying to the gluon propagator gives
\begin{equation}
\frac{d}{dn_\beta}N^{\nu\nu'}=
-\frac{1}{n\cdot l}(l^\nu N^{\beta\nu'}+l^{\nu'} N^{\nu\beta})\;.
\label{dgp}
\end{equation}
The loop momentum $l^\nu$ ($l^{\nu'}$)
contracts with a vertex in $F$, which is then replaced by a special vertex 
\begin{eqnarray}
{\hat v}_\beta=\frac{n^2v_\beta}{v\cdot nn\cdot l}\;.
\label{va}
\end{eqnarray}
This special vertex can be read off the combination of
Eqs.~(\ref{cph}) and (\ref{dgp}).

The contraction of $l^\nu$ hints the application of the Ward
identities \cite{L2,L3},
\begin{equation}
\frac{i(\not k+\not l)}{(k+l)^2}(-i\not l)\frac{i\not k}{k^2}
=\frac{i\not k}{k^2}-\frac{i(\not k+\not l)}{(k+l)^2}\;,
\end{equation}
for the quark-gluon vertex, and
\begin{equation}
l^\nu\frac{-iN^{\alpha\mu}(k+l)}{(k+l)^2}\Gamma_{\mu\nu\lambda}
\frac{-iN^{\lambda\gamma}(k)}{k^2}
=-i\left[\frac{-iN^{\alpha\gamma}(k)}{k^2}-
\frac{-iN^{\alpha\gamma}(k+l)}{(k+l)^2}\right]\;,
\label{ward}
\end{equation}
for the triple-gluon vertex $\Gamma_{\mu\nu\lambda}$. Similar identities
for other types of vertices can be derived easily.
Summing all the diagrams with different differentiated gluons, those
embedding the special vertices cancel by pairs, leaving the one where 
the special vertex moves to the outer end of the parton line \cite{CS}.
We obtain the formula,
\begin{equation}
-x\frac{d}{dx}F(x,p_T)=2{\bar F}(x,p_T)\;,
\label{df}
\end{equation}
described by Fig.~1(a), where the new function $\bar F$ contains one
special vertex represented by a square. The coefficient 2 comes from the
equality of the new functions with the special vertex on either of the two
parton lines. Note that Eq.~(\ref{df}) is an exact
consequence of the Ward identity without approximation \cite{CS}. An
approximation will be introduced, when ${\bar F}$ is related to $F$ by
factorizing out the subdiagram containing the special vertex, such that
Eq.~(\ref{df}) reduces to a differential equation of $F$.

It is known that factorization holds in the leading regions.
The leading regions of the loop momentum $l$ flowing through the special 
vertex are soft and hard, since the factor $1/n\cdot l$ with $n^2\not=0$ in
Eq.~(\ref{va}) suppresses collinear divergences \cite{CS}. For soft and
hard $l$, the subdiagram containing the special vertex is factorized
according to Figs.~1(b) and 1(c), respectively. Fig.~1(b) collects the soft
divergences of the subdiagram by eikonalizing the gluon propagator, which
will be explained below. We extract the color factor from the relation
$f_{abc}f_{bdc}=-N_c\delta_{ad}$, the indices $a,b,\dots$ being
indicated in Fig.~1(b), and $N_c=3$ the number of colors. The
corresponding factorization formula is then written as
\begin{eqnarray}
{\bar F}_s(x,p_T)&=&
iN_cg^2\int\frac{d^{4}l}{(2\pi)^4}
\Gamma_{\mu\nu\lambda}{\hat v}_\beta
[-iN^{\nu\beta}(l)]
\frac{-iN^{\lambda\gamma}(xP)}{-2xP\cdot l}
\nonumber \\
& &\times\left[2\pi i\delta(l^2)F(x+l^+/P^+,|{\bf p}_T+{\bf l}_T|)
+\frac{\theta(p_T^2-l_T^2)}{l^2}F(x,p_T)\right],
\nonumber\\
& &
\label{kf}
\end{eqnarray}
where $iN_c$ comes from the product of the overall coefficient $-i$ in
Eq.~(\ref{ward}) and the color factor $-N_c$ extracted above, $g$ is the
coupling constant, and the triple-gluon vertex for vanishing $l$ is given by
\begin{equation}
\Gamma_{\mu\nu\lambda}=
-g_{\mu\nu}xP_{\lambda}-g_{\nu\lambda}xP_{\mu}+2g_{\lambda\mu}xP_{\nu}\;.
\label{tri}
\end{equation}
The denominator $-2xP\cdot l$ comes from the eikonal
approximation $(xP-l)^2\approx -2xP\cdot l$. The first term in the brackets
corresponds to the real gluon emission, where
$F(x+l^+/P^+,|{\bf p}_T+{\bf l}_T|)$ implies that the parton coming out of
the proton carries the momentum components $xP^++l^+$ and
${\bf p}_T+{\bf l}_T$ in order to radiate a real gluon of momentum
$l$. The second term corresponds to the virtual gluon emission,
where the $\theta$ function sets the upper bound of $l_T$ to $p_T$ to ensure
a soft momentum flow.

It can be easily shown that the contraction of $P$ with a vertex in the
quark box diagram the partons attach, or with a vertex in the gluon
distribution function, gives a contribution down by a power $1/s$,
$s=(P+q)^2$, compared to the contribution from the contraction with
${\hat v}_\beta$. Following this observation, Eq.~(\ref{kf}) is
reexpressed as
\begin{eqnarray}
{\bar F}_s(x,p_T)&=&
iN_cg^2\int\frac{d^{4}l}{(2\pi)^4}N^{\nu\beta}(l)
\frac{{\hat v}_\beta v_\nu}{v\cdot l}
\left[2\pi i\delta(l^2)F(x+l^+/P^+,|{\bf p}_T+{\bf l}_T|)\right.
\nonumber \\
& &\left.
+\frac{\theta(p_T^2-l_T^2)}{l^2}F(x,p_T)\right]\;.
\label{kf1}
\end{eqnarray}
The eikonal vertex $v_\nu$ comes from the term $xP_\nu$ (divided by $xP^+$)
in Eq.~(\ref{tri}), and the eikonal propagator $1/v\cdot l$ from
$1/(xP\cdot l)$, which is represented by a double line in Fig.~1(b).
The remaining metric tensor $g^{\mu\gamma}$ has been absorbed into $F$.
We apply the strong rapidity ordering $x+l^+/P^+\gg x$ to the
evaluation of Eq.~(\ref{kf1}), that is, approximate
$F(x+l^+/P^+,|{\bf p}_T+{\bf l}_T|)$ by its dominant value
$F(x,|{\bf p}_T+{\bf l}_T|)$. The integration over $l^-$ and $l^+$ to
infinity gives
\begin{eqnarray}
{\bar F}_s(x,k_T)&=&\frac{{\bar \alpha}_s}{2}
\int\frac{d^{2}l_T}{\pi}
\frac{-n^2}{2n^-[n^+l_T^2+2n^-l^{+2}]}|^{l^+=\infty}_{l^+=0}
\nonumber\\
& &\times\left[F(x,|{\bf k}_T+{\bf l}_T|)
-\theta(k_T^2-l_T^2)F(x,k_T)\right]\;,
\nonumber\\
&=&\frac{{\bar \alpha}_s}{2}
\int\frac{d^{2}l_T}{\pi l_T^2}
\left[F(x,|{\bf k}_T+{\bf l}_T|)-\theta(k_T^2-l_T^2)F(x,k_T)\right]\;,
\label{kf2}
\end{eqnarray}
with ${\bar \alpha}_s=N_c\alpha_s/\pi$ and $n=(n^+,n^-,{\bf 0})$ has been
assumed for convenience. The first line of the above
formulas demonstrates explicitly how the $n$ dependence cancels in the
evaluation of Fig.~1(b).

It can be shown that the contribution from the first diagram of Fig.~1(c)
vanishes like \cite{L3}
\begin{eqnarray}
& &-\frac{{\bar\alpha}_s}{2}\int\frac{d^{2}l_T}{\pi}
\left[\frac{1}{l_T^2}-\frac{1}{l_T^2+(xP^+\nu)^2}\right.
\nonumber \\
& &\left.
-\frac{1}{2}\frac{xP^+\nu}{[l_T^2+(xP^+\nu)^2]^{3/2}}
\ln\frac{\sqrt{l_T^2+(xP^+\nu)^2}-xP^+\nu}
{\sqrt{l_T^2+(xP^+\nu)^2}+xP^+\nu}\right]
\label{gpp}
\end{eqnarray}
at $x\to 0$, with the factor $\nu=\sqrt{(v\cdot n)^2/n^2}$. This is the
reason $F$ does not acquire an explicit dependence on the large scale $P^+$
as stated before, and the transverse degrees of freedom must be taken into
account, leading to the $p_T$-factorization theorem \cite{J}. The
$n$ dependence residing in Fig.~1(c) then disappears with the vanishing of
Eq.~(\ref{gpp}).
If neglecting the contribution from Fig.~1(c), {\it i.e.}, adopting
${\bar F}={\bar F}_s$, Eq.~(\ref{df}) becomes
\begin{eqnarray}
\frac{dF(x,p_T)}{d\ln(1/x)}=
{\bar \alpha}_s(p_T)\int\frac{d^{2}l_T}{\pi l_T^2}
\left[F(x,|{\bf p}_T+{\bf l}_T|)-\theta(p_T^2-l_T^2)F(x,p_T)\right]\;,
\label{bfkl}
\end{eqnarray}
which is exactly the BFKL equation. The argument of ${\bar \alpha}_s$ has
been set to the natural scale $p_T$. It is then understood that the
subdiagram containing the special vertex plays the role of the evolution 
kernel. Obviously, the BFKL kernel is gauge invariant after deriving the
evolution equation. Now we make the vector $n$ approach
$\delta^{\mu-}$, and the definition of $F$ returns to the standard one.

A careful examination reveals that it is not proper to extend the loop
momentum $l^+$ of a real gluon to infinity when deriving Eq.~(\ref{kf2}),
since the behavior of $F(x+l^+/P^+)$, vanishing at the large momentum
fraction, should introduce an upper bound of $l^+$. To obtain a more
reasonable evolution kernel, we propose to truncate $l^+$ at
some scale, and a plausible choice of this scale is of order $Q$. The cutoff
of $l^+$ introduces a $n$ dependence. However, this $n$ dependence can be
absorbed into the cutoff, which will be regarded as a fitting parameter and
determined by the data of $F_2$ (for some value of $Q$). Then the gauge
dependence does not appear explicitly in the modified evolution equation.
It can be verified that the predictions are insensitive to the cutoff,
and a good choice of the cutoff is $Q/(2\nu)$, where $\nu$ is the gauge
factor appearing in Eq.~(\ref{gpp}). Performing the integration over $l^-$
and $l^+$, Eq.~(\ref{kf2}) becomes
\begin{eqnarray}
{\bar F}_s(x,p_T,Q)
&=&\frac{{\bar\alpha}_s}{2}\int\frac{d^{2}l_T}{\pi}
\frac{-1}{l_T^{2}+4\nu^2l^{+2}}|_{l^+=0}^{l^+=Q/(2\nu)}
\nonumber \\
& &\times\left[F(x,|{\bf p}_T+{\bf l}_T|,Q)
-\theta(p_T^2-l_T^2)F(x,p_T,Q)\right]\;,
\label{kf0}
\end{eqnarray}
where the $Q$ dependence of $F$ from the constraint of phase space for the
real gluon emission has been indicated. With this modification,
the real gluon has smaller phase space at lower $Q$.

We derive the modified BFKL equation
\begin{eqnarray}
\frac{dF(x,p_T,Q)}{d\ln(1/x)}
&=&{\bar\alpha}_s(p_T)\int\frac{d^{2}l_T}{\pi l_T^2}
[F(x,|{\bf p}_T+{\bf l}_T|,Q)-\theta(p_T^2-l_T^2)F(x,p_T,Q)]
\nonumber\\
& &-{\bar\alpha}_s(p_T)\int\frac{d^{2}l_T}{\pi}
\frac{F(x,|{\bf p}_T+{\bf l}_T|,Q)}{l_T^2+Q^{2}}\;,
\label{mbf}
\end{eqnarray}
where the first and last terms correspond to the lower and upper bounds of 
$l^+$, respectively. Obviously, Eq.~(\ref{mbf}) approaches the conventional
equation (\ref{bfkl}) in the $Q\to \infty$ limit. The first term on
the right-hand side of Eq.~(\ref{mbf}) is responsible for the rise of $F$
at small $x$, and the last term acts to moderate the rise \cite{L3}. We then
expect that the ascent of $F$ is slower at smaller $Q$, for which the effect
of the last term is stronger, and deviates from that predicted by the
conventional BFKL equation. Note that the $Q$ dependence is attributed to
the constraint of the phase space for radiative corrections, instead of to
the $\ln Q$ summation.

\vskip 1.0cm

\centerline{\large\bf III. EVOLUTION IN $x$}
\vskip 0.5cm

Before proceeding with the numerical analysis, we extract the behavior
of the gluon distribution function at small $x$ from the modified
BFKL equation, and compare it with those from the DGLAP, BFKL, and CCFM
equations. It will be observed that the modified BFKL equation and
the CCFM equation give similar evolution in $x$. It is most convenient
to consider the Fourier transformed version of the equations.
We reexpress Eq.~(\ref{mbf}) as \cite{L3}
\begin{eqnarray}
\frac{dF(x,p_T,Q)}{d\ln(1/x)}
&=&{\bar\alpha}_s\int\frac{d^{2}l_T}{\pi l_T^2}
[F(x,|{\bf p}_T+{\bf l}_T|,Q)-\theta(Q_0^2-l_T^2)F(x,p_T,Q)]
\nonumber\\
& &-{\bar\alpha}_s\int\frac{d^{2}l_T}{\pi}
\frac{F(x,|{\bf p}_T+{\bf l}_T|,Q)}{l_T^2+Q^{2}}\;.
\label{mbfb}
\end{eqnarray}
with a fixed coupling constant ${\bar\alpha_s}$. Note that
the loop momentum $l_T$ in the virtual gluon emission is truncated at the
scale $Q_0$, instead of at $p_T$. This modification is acceptable, 
since the virtual gluon contribution plays the role of a soft regulator for 
the real gluon emission, and setting the cutoff to $Q_0$ serves the 
same purpose. Furthermore, the replacement of $p_T$ by $Q_0$ allows us to 
solve Eq.~(\ref{mbfb}) analytically, and the solution of $F$ maintains 
the essential BFKL features. 

The Fourier transform of Eq.~(\ref{mbfb}) to the conjugate $b$ space
leads to
\begin{eqnarray}
\frac{d{\tilde F}(x,b,Q)}{d\ln(1/x)}&=&-S(b,Q){\tilde F}(x,b,Q)\;,
\label{bfb}
\end{eqnarray}
with 
\begin{equation}
S(b,Q)=2{\bar\alpha}_s[\ln(Q_0 b)+\gamma-\ln 2+K_0(Qb)]\;.
\label{es}
\end{equation}
where the Bessel function $K_0$ comes from the last term on the right-hand
side of Eq.~(\ref{mbfb}), and $\gamma$ is the Euler constant.
Equation (\ref{bfb}) is trivially solved to give
\begin{eqnarray}
{\tilde F}(x,b,Q)\propto \exp[-S(b,Q)\ln(x_0/x)]\;,
\label{sbfb}
\end{eqnarray}
$x_0$ being the initial momentum fraction below which the gluon
distribution function begins to evolve according to the BFKL equation
Transforming Eq.~(\ref{sbfb}) back to momentum space, we have
\begin{eqnarray}
F(x,p_T,Q)&=&\int_0^\infty bdb J_0(p_Tb){\tilde F}(x,b,Q)\;.
\label{sbf}
\end{eqnarray}
Assuming that the above integral is dominated by the contribution from
the small $b$ region, $F$ is expected to grow as
\begin{equation}
{\tilde F}(x,0,Q)\propto
\exp[2{\bar \alpha}_s\ln(x_0/x)\ln(Q/Q_0)]\;.
\label{bfev}
\end{equation}
This power-law rise with $1/x$ is steeper compared to the DGLAP evolution,
\begin{equation}
\exp\left[4\sqrt{\frac{N_c}{\beta_0}\ln\frac{1}{x}
\ln\frac{\ln(Q/\Lambda_{\rm QCD})}{\ln(Q_0/\Lambda_{\rm QCD})}}\right]\;,
\end{equation}
$\beta_0$ being the first coefficient of the QCD beta function, and is
more sensitive to the variation of $Q$ compared to the conventional BFKL
evolution,
\begin{equation}
\exp[4\ln 2{\bar \alpha}_s\ln(x_0/x)]\;.
\label{coev}
\end{equation}
Remind that the DGLAP evolution with $x$ is not strong enough, while the
conventional BFKL rise is $Q$-independent. Equations
(\ref{bfev})-(\ref{coev}) then make clear why the modified BFKL equation
can explain the data well.

At the same time, we shall study the CCFM equation, whose $Q$ dependence
comes from the $\ln Q$ summation. It is written as
\begin{eqnarray}
F(x,p_T,Q)&=&F^{(0)}(x,p_T,Q)+\int_x^1 dz
\int\frac{d^2q}{\pi q^2}\theta(Q-zq)
\Delta_S(Q,zq)
\nonumber \\ 
& &\hspace{2.0cm}\times
{\tilde P}(z,q,p_T)F(x/z,|{\bf p}_T+(1-z){\bf q}|,q)\;,
\label{ccfm}
\end{eqnarray}
with the splitting function 
\begin{equation}
{\tilde P}={\bar\alpha_s}(p_T)
\left[\frac{1}{1-z}+\Delta_{NS}(z,q,p_T)\frac{1}{z}+z(1-z)\right]\;.
\label{pgg}
\end{equation}
The so-called ``Sudakov" exponential $\Delta_S$ and the ``non-Sudakov" 
exponentials $\Delta$ are given by
\begin{eqnarray}
\Delta_S(Q,zq)&=&\exp\left[-{\bar\alpha_s}
\int_{(zq)^2}^{Q^2}\frac{dp^2}{p^2}
\int_{0}^{1-p_T/p}\frac{dz'}{1-z'}\right]
\nonumber \\
\Delta_{NS}(z,q,p_T)&=&
\exp\left[-{\bar\alpha_s}\int_{z}^{z_0}\frac{dz'}{z'}
\int_{(z'q)^2}^{p_T^2}\frac{dp^2}{p^2}\right]\;.
\label{nons}
\end{eqnarray}
The upper bound $z_0$ of the variable $z'$ in Eq.~(\ref{nons}) takes the 
values \cite{CCFM,KMS1},
\begin{eqnarray}
   & & 1 \hspace{1.5cm}  {\rm if}\;\; 1\le (p_T/q)
\nonumber \\
z_0&=& p_T/q \hspace{1.0cm}  {\rm if}\;\; z < (p_T/q) < 1
\nonumber \\
   & & z \hspace{1.5cm}   {\rm if}\;\; (p_T/q)\le z\;.
\end{eqnarray}   
The Sudakov exponential $\Delta_S$ collects the contributions from the
rung gluons obeying the strong angular ordering, and is the result of
the $\ln Q$ summation. Those gluons which do not obey the angular
ordering are grouped into $\Delta_{NS}$.
For the simpler derivation of the CCFM equation using the resummation
technique, refer to \cite{L3}.
In the small $x$ region we adopt the approximate form of the 
CCFM equation \cite{KMS1},
\begin{eqnarray}
F(x,p_T,Q)&=&F^{(0)}(x,p_T,Q)+{\bar \alpha}_s(p_T)\int_x^1 \frac{dz}{z}
\int\frac{d^2q}{\pi q^2}\theta(Q-zq)\theta(q-\mu)
\nonumber \\ 
& &\hspace{1.5cm}\times
\Delta_{NS}(z,q,p_T)
F(x/z,|{\bf p}_T+(1-z){\bf q}|,q)\;.
\label{cca}
\end{eqnarray}
The extra theta function $\theta(q-\mu)$ introduces an infrared cutoff of
the variable $q$. The flat nonperturbative driving term $F^{(0)}$ will be
defined later.

We extract the approximate CCFM evolution in $x$ as a comparision.
Applying $d/d\ln(1/x)$ to both sides of Eq.~(\ref{cca}) gives
\begin{eqnarray}
\frac{dF(x,p_T,Q)}{d\ln(1/x)}&\approx
&{\bar \alpha}_s\int_x^1 \frac{dz}{z}
\int\frac{d^2q}{\pi q^2}\theta(Q-zq)\theta(q-\mu)\Delta_{NS}(z,q,p_T)
\nonumber \\ 
& &\hspace{1.5cm}\times
\frac{d}{d\ln(1/x)}F(x/z,|{\bf p}_T+{\bf q}|,Q)\;.
\end{eqnarray}
Again, we assume a fixed coupling constant ${\bar\alpha}_s$.
The derivative of the flat driving term $F^{(0)}$ has
been dropped. The differentiation of the lower bound of $z$ vanishes,
because of $F(\xi=1)=0$. The arguments $|{\bf p}_T+(1-z){\bf q}|$ and $q$
of $F$ in the integrand have been replaced by
$|{\bf p}_T+{\bf q}|$ and $Q$, respectively, for simplicity. The identity
$dF/d\ln(1/x)=dF/d\ln z$ and the neglect of the derivatives of other
factors in the integrand lead to
\begin{eqnarray}
\frac{dF(x,p_T,Q)}{d\ln(1/x)}&\approx&{\bar \alpha}_s
\int_x^1 dz \frac{d}{dz}\int\frac{d^2q}{\pi q^2}\theta(Q-zq)\theta(q-\mu)
\nonumber \\ 
& &\hspace{1.5cm}\times
\Delta_{NS}(z,q,p_T)F(x/z,|{\bf p}_T+{\bf q}|,Q)\;,
\nonumber \\
&=&{\bar \alpha}_s\int\frac{d^2q}{\pi q^2}\theta(Q-q)\theta(q-\mu)
F(x,|{\bf p}_T+{\bf q}|,Q)\;.
\end{eqnarray}
To derive the last line of the above formulas, we have employed
$F(\xi=1)=0$ and $\Delta_{NS}(1,q,p_T)=1$ according to Eq.~(\ref{nons}).
The Fourier transform of the above equation is given by
\begin{eqnarray}
\frac{d{\tilde F}}{d\ln(1/x)}\approx 2{\bar \alpha}_s\ln(Q/\mu){\tilde F}
\end{eqnarray}
in the small $b$ region. It is obvious that the CCFM equation predicts
a similar evolution in $x$ to Eq.~(\ref{bfev}). This similarity will
be verified numerically in the next section. However, its $Q$ dependence
is due to the transverse momentum ordering as indicated by the function
$\theta(Q-q)$, a source different from the constraint on the longitudinal
momenta of real gluons in the modified BFKL equation.

Another interesting observation is that the CCFM equation gives a steeper
rise at small $x$ than the DGLAP equation does. The reason is attributed to
the different choices of the argument of the running coupling constant
in the splitting function, which is $p_T$ in the former
(see Eq.~(\ref{pgg})) and $q$ in the latter. Hence, $\alpha_s(p_T)$ does
not run in fact, as the variable $q$ is integrated over in the CCFM equation
(\ref{cca}). While $\alpha_s(q)$ evolves to $\alpha_s(Q)$ when solving the
DGLAP equation. Because of $\alpha_s(p_T) > \alpha_s(Q)$, the CCFM evolution
is stronger.

\vskip 1.0cm

\centerline{\large\bf IV. NUMERICAL RESULTS}
\vskip 0.5cm

We solve both the modified and conventional BFKL equations
numerically following the repetation method proposed in \cite{KMS1}
by starting with a ``flat" gluon distribution function \cite{KMS1,CG},
\begin{equation}
F^{(i)}(x,p_T)=\frac{3}{Q_0^2}N_g(1-x)^5\exp(-p_T^2/Q_0^2)\;,
\label{fi}
\end{equation}
with $Q_0=1$ GeV. The active flavor number $n_f$ in the running coupling
constant $\alpha_s$ is set to 4. The normalization constant $N_g$ will be
determined by the data of the structure function $F_2(x,Q^2)$ at some $Q$,
and then employed to make predictions for other $Q$. We solve
Eqs.~(\ref{bfkl}) and (\ref{mbf}) with $F^{(i)}$ in Eq.~(\ref{fi}) inserted
into their right-hand sides. Substitute the obtained solution of $F$
into the right-hand sides of Eqs.~(\ref{bfkl}) and (\ref{mbf}), and solve
them again. Repeat this procedure until a limit solution of $F$ is reached.
In the numerical analysis we introduce a lower bound $p_T >1$ GeV to avoid
the infrared divergence in ${\bar \alpha}_s(p_T)$ from the $p_T$
diffusion at small $x$. The CCFM equation is also solved in a similar way
with the driving term $F^{(0)}$ 
\begin{eqnarray}
F^{(0)}(x,p_T,Q)&=&\frac{3}{Q_0^2}N_g\exp(-p_T^2/Q_0^2)
\int_x^1 \frac{dz}{z}\theta(Q-zp_T)
\nonumber \\ 
& &\times \Delta_{NS}(z,p_T,p_T)\frac{d(1-x/z)^5}{d\ln(z/x)}\;,
\label{cci}
\end{eqnarray}
wich reduces to $F^{(i)}$, when the functions $\theta$ and $\Delta_{NS}$
are set to unity.

The expressions of the structure functions $F_{2,L}$,
according to the $p_T$-factorization theorem \cite{J}, are given by
\begin{eqnarray}
F_2(x,Q^2)&=&\int_x^1 \frac{d\xi}{\xi}\int_0^{p_c} \frac{d^2p_T}{\pi}
H_2(x/\xi,p_T,Q)F(\xi,p_T,Q)+\frac{3}{2}F_L(x,Q^2),
\nonumber\\
& &\label{f2}\\
F_L(x,Q^2)&=&\int_x^1 \frac{d\xi}{\xi}\int_0^{p'_c} \frac{d^2p_T}{\pi}
H_L(x/\xi,p_T,Q)F(\xi,p_T,Q),
\label{fl}
\end{eqnarray}
$p_c$ and $p'_c$ being the upper bounds of $p_T$ which will be specified
later. The hard scattering subamplitudes $H_{2,L}$ denote the
contributions from the quark box diagrams in Fig.~2, where both the
incoming photon and gluon are off shell by $q^2=-Q^2$ and $p^2=-p_T^2$,
respectively, $p=(\xi P^+,0,{\bf p}_T)$ being the parton momentum and
$\xi$ the momentum fraction. They are written as               
\begin{eqnarray}
H_2&=&e_q^2\frac{\alpha_s}{\pi}Cz
\Biggl\{\left[z^2+(1-z)^2-2z(1-2z)\frac{p_T^2}{Q^2}+2z^2\frac{p_T^4}{Q^4}
\right]
\nonumber \\
& &\times\frac{1}{A}\ln\frac{1+A}{1-A}-2\Biggr\}\;,
\label{hgc}
\end{eqnarray}
and $H_L=H_L^{a+b}+H_L^c$, with
\begin{eqnarray}
H_L^{a+b}&=&-2e_q^2\frac{\alpha_s}{\pi}Cz
\Biggl\{\frac{1}{2}-\frac{x}{z}
+z\frac{p_T^2}{Q^2}\frac{B^2}{A^2}
\nonumber \\
& &+\frac{2z^2}{A^2}
\left[\left(1-2z\frac{p_T^2}{Q^2}\right)^2-2z\frac{p_T^2}{Q^2}
B^2\right]
\nonumber \\
& &\times\left(\frac{p_T^2}{Q^2}-\frac{p_T^2}{Q^2}\frac{1}{A}
\ln\frac{1+A}{1-A}
+\frac{1}{4z^2}\right)\Biggr\}
\label{ppab}
\end{eqnarray}
the sum of the contributions from Figs.~2(a) and 2(b), and
\begin{eqnarray}
H_L^c&=&-2e_q^2\frac{\alpha_s}{\pi}Cz
\Biggl\{\left[\frac{x^2}{z}-2x^2\frac{p_T^2}{Q^2}\right.
\nonumber \\
& &\left.-\left(1-\frac{p_T^2}{Q^2}\right)\left(2zx-z-2z^2\frac{p_T^2}{Q^2}
\frac{B^2}{A^2}\right)\right]\frac{1}{A}
\ln\frac{1+A}{1-A}
\nonumber \\
& &-\left[(z-zx)\left(1-\frac{p_T^2}{Q^2}\right)
\left(1-2z\frac{p_T^2}{Q^2}\right)-xA^2\right]\frac{1}{A^2}
\ln\frac{1+A^2}{1-A^2}
\nonumber \\
& &-\left[2x\left(1-2z\frac{p_T^2}{Q^2}\right)\right.
\nonumber \\
& &\left.-z\left(1-\frac{p_T^2}{Q^2}\right)
\frac{1-2z(3-z)p_T^2/Q^2+6z^2p_T^4/Q^4}{A^2}\right]
\nonumber \\
& &\times\frac{1}{A^2}
\left(\frac{1}{A}\ln\frac{1+A}{1-A}
-2\right)\Biggr\}
\label{ppc}
\end{eqnarray}
from Fig.~2(c). The factors $A$, $B$ and $C$ are $A=\sqrt{1-4z^2p_T^2/Q^2}$, 
$B=\sqrt{1-z-z p_T^2/Q^2}$ and $C=1/2$. The detailed derivation of $H_{2,L}$
is presented in the Appendix. We have assumed a vanishing charm quark mass
for simplicity. To require $H_{2,L}$ be meaningful, the upper bounds of 
$p_T$ are set to
\begin{equation}
p_c=\min\left(Q,\frac{\xi}{2x}Q\right)\;,\;\;\;\;
p'_c=\min\left(Q,\frac{\xi-x}{x}Q\right)\;.
\end{equation}
It is easy to observe that the terms in the braces of $H_2$ approaches the 
splitting function 
\begin{equation}
P_{qg}(z)=C[z^2+(1-z)^2] 
\end{equation}
in the $p_T\to 0$ limit.

We evaluate Eq.~(\ref{f2}) for $Q^2=15$ GeV$^2$ with
$F(\xi,p_T,Q)$ obtained from the modified BFKL equation, and then determine
the normalization constant as $N_g=3.29$ from the data fitting.
When $Q^2$ varies, we adjust $N_g$ such that $xg$ has a fixed normalization
$\int_0^1 xgdx$. $F_2$ for $Q^2=8.5$, 12, 25, and 50 GeV$^2$ are
then computed, and results along with the HERA data \cite{H1} are displayed 
in Fig.~3. It is obvious that our predictions agree with the data well. The 
curves have a slower rise at smaller $Q$, which is the consequence of the 
phase space constraint for real gluon emissions.
For comparision, we show the results from the
conventional BFKL equation, and from the CCFM equation. The former are
almost $Q$-independent, and close to those from the modified BFKL equation
only at large $Q^2=50$ GeV$^2$. Therefore, their match with the data is not
very satisfactory, especially in the low $Q$ region. The latter, also
consistent with the data, are similar to those from the modified BFKL
equation as expected.

The normalization constants $N_g$ for $F_2$ are then employed to evaluate
the structure function $F_L(x,Q^2)$ using Eq.~(\ref{fl}) for the
corresponding $Q$. The predictions for $Q^2=8.5$, 12, 15, 25, and 50 GeV$^2$
are presented in Fig.~4, which behave in a similar way to $F_2(x,Q^2)$, but
are smaller in magnitude. Note that the uncertainty of the experimental
data \cite{H12} shown in Fig.~4 is still large: The vertical lines adjacent
to $x=10^{-4}$ for $Q^2=8.5$ and 12 GeV$^2$ represent the error bars. The
central values slightly above 0.5 are outside of the plot range. Hence, it
is not difficult to explain the current data. To draw a concrete conclusion,
more precise data are necessary.

The gluon density $xg$ is defined by the integral of the unintegrated 
gluon distribution function over $p_T$ \cite{KMS1},
\begin{eqnarray}
xg(x,Q^2)=\int_0^Q\frac{d^2 p_T}{\pi}F(x,p_T,Q)\;.
\label{gbf}
\end{eqnarray}
The dependnece of $xg$ on $x$ for $Q^2=8.5$, 15, and 50 GeV$^2$ derived
from the modified BFKL equation is displayed in Fig.~5. If letting
the parameter $\lambda$ characterize the behavior of $xg\sim x^{-\lambda}$ 
at small $x$, we find the values $\lambda\approx 0.3$ for 
$Q^2=8.5$ GeV$^2$ and $\lambda\approx 0.4$ for $Q^2=50$ GeV$^2$. The former 
is consistent with that obtained from a phenomenological fit to the HERA
data \cite{MSR}. The latter is close to that from solving the conventional
BFKL equation numerically \cite{AKMS}, {\it i.e.}, from hard pomeron
contributions. As explained before, the smaller $\lambda$ at lower $Q$
is due to the more restricted phase space for real gluon emissions.
The flat gluon density corresponding to
soft pomeron exchanges can then be understood following this vein.

\vskip 1.0cm
\centerline{\large\bf V. CONCLUSION}
\vskip 0.5cm

In this paper we have studied in details the modified BFKL equation
proposed from the viewpoint of the resummation technique. This equation
possesses a $Q$ dependence from the constraint of the phase space for soft
real gluon emissions. It gives rise to the desired behaviors in $x$ and in
$Q$ of the gluon distribution function and the structure functions
$F_{2,L}$, which are consistent with the HERA data. The predictions 
are close to those from the CCFM equation, whose $Q$ dependence is, however,
due to the $\ln Q$ summation. The rise of the gluon density at small $x$
predicted by the modified BFKL equation is steeper than that by the DGLAP
equation, and is more sensitive to the variation of $Q$ compared to that
by the conventional BFKL equation. We have also explained why the
conventional BFKL equation overestimates the real gluon contributions
and thus the rise of the structure functions
in the low $Q$ region. Real gluons, responsible for the BFKL rise,
in fact should have more restricted phase space at smaller $Q$.

This work demonstrates the simplicity and success of the resummation
technique in the study of the DIS structure functions at small $x$. However,
the power-law rise of the  structure functions predicted by either the
conventional or modified BFKL equation violates the unitarity bound \cite{F}.
A further application of our formalism to the issue of the unitarity of the
BFKL evolution will be published elsewhere \cite{LU}.

\vskip 1.0cm
This work was supported by the National Science Council of Republic of
China under the Grant No. NSC87-2112-M-006-018.

\vskip 2.0cm

\centerline{\large \bf Appendix} 
\vskip 0.5cm

In this Appendix we extract the hard scattering subamplitudes from the 
box diagrams in Fig.~2. The photon and gluon with the off-shell 
momenta $q^2=-Q^2$ and $p^2=-p_T^2$, respectively, annihilate into a quark 
with momentum $k$ and an antiquark with $k'$. These diagrams can be regarded 
as a parton-level cross section, which are then convoluted with the
unintegrated gluon distribution function $F$ as shown in Eq.~(\ref{f2}).
The structure functions $F_{2,L}$ are extracted
from the DIS hadronic tensor $W_{\mu\nu}$ through the contractions
\begin{eqnarray}
-g^{\mu\nu}W_{\mu\nu}=\frac{F_2}{x}-\frac{3}{2x}F_L\;,
\nonumber \\
P^\mu P^\nu W_{\mu\nu}=\frac{Q^2}{4x^2}\frac{F_L}{2x}\;,
\end{eqnarray}
where the longitudinal structure function $F_L$ is, in terms of the 
standard ones $F_{1,2}$, written as $F_L=F_2-2xF_1$.

In the infinite-momentum frame of the proton, the only nonvanishing
component of $P$ is $P^+$, and the gluon carries the parton momentum
$\xi P^++{\bf p}_T$. In the center-of-mass frame of the photon and gluon,
the gluon momentum is transformed into $p$ with the space component
${\bf p}$ in the $z$ direction. The relation ${\bf q}+{\bf p}=0$ leads to 
\begin{equation}
q_0^2=q^2+|{\bf q}|^2=-Q^2+|{\bf p}|^2=p_0^2-p^2-Q^2\;.
\label{q0}
\end{equation}
>From the parton-level Bjorken variable
\begin{equation}
z=\frac{x}{\xi}=\frac{Q^2}{2p\cdot q}=
\frac{Q^2}{2(p_0q_0+|{\bf p}||{\bf q}|)}\;,
\end{equation}
with Eq.~(\ref{q0}) inserted, we solve for $p_0$, which is expressed as
\begin{equation}
p_0=\frac{Q}{2}\frac{1+2z p^2/Q^2}{\sqrt{z(1-z+z p^2/Q^2)}}\;.
\end{equation}
It is then easy to obtain
\begin{eqnarray}
|{\bf p}|&=&\frac{Q}{2}\frac{\sqrt{1+4z^2p^2/Q^2}}
{\sqrt{z(1-z+z p^2/Q^2)}}\;,
\\
q_0&=&\frac{Q}{2}\frac{1-2z}{\sqrt{z(1-z+z p^2/Q^2)}}\;.
\end{eqnarray}
Since each of the quarks shares half of the total energy, we have
\begin{eqnarray}
k_0=k'_0=\frac{1}{2}(p_0+q_0)=\frac{Q}{2}\frac{1-z+z p^2/Q^2}
{\sqrt{z(1-z+z p^2/Q^2)}}\;.
\end{eqnarray}
Similarly, the invariants $s=(p+q)^2$, $t=(p-k')^2$ and $u=(p-k)^2$
are written as
\begin{eqnarray}
s&=&\frac{Q^2}{x}\left(1-z+z\frac{p^2}{Q^2}\right)\;,
\nonumber \\
t&=&-\frac{Q^2}{2z}\left(1+\sqrt{1+4z^2\frac{p^2}{Q^2}}\cos\theta
\right)\;,
\nonumber \\
u&=&-\frac{Q^2}{2z}\left(1-\sqrt{1+4z^2\frac{p^2}{Q^2}}\cos\theta
\right)\;,
\end{eqnarray}
where $\theta$ is the angle between the momenta ${\bf k}$ and ${\bf p}$.

We first compute $-g^{\mu\nu}{\hat W}_{\mu\nu}$, where ${\hat W}_{\mu\nu}$
is the parton-level hadronic tensor. Contracting $-g^{\mu\nu}$ to the
box diagrams, the squared amplitude from Fig.~2(a) is given by
\begin{eqnarray}
(-g^{\mu\nu}{\hat W}_{\mu\nu})_a
=\frac{1}{4}e_q^2g^2C
Tr\left[\not k'\gamma_\nu\frac{\not k-\not q}{(k-q)^2}\gamma_\mu
\not k\gamma^\mu\frac{\not k-\not q}{(k-q)^2}\gamma^\nu\right]\;,
\end{eqnarray}
where the factor $1/4$ denotes the average over the spins of the 
photon and gluon, $C=1/2$ is the color factor from the identity
$Tr(T^aT^b)=(1/2)\delta^{ab}$, $T^{a,b}$ being the color matrices,
and $e_q$ is the charge of the quark $q$ in unit of $e$. A  
straightforward calculation leads to
\begin{eqnarray}
(-g^{\mu\nu}{\hat W}_{\mu\nu})_a
=8\pi e_q^2\alpha_sC\left(\frac{tu+p^2Q^2}{t^2}
\right)\;.
\end{eqnarray}
Since Fig.~2(b) is idential to Fig.~2(a) under the interchange of $k$ and 
$k'$, we have $(-g^{\mu\nu}{\hat W}_{\mu\nu})_b=
(-g^{\mu\nu}{\hat W}_{\mu\nu})_a(t\leftrightarrow u)$.
A similar evaluation of Fig.~2(c) gives
\begin{eqnarray}
(-g^{\mu\nu}{\hat W}_{\mu\nu})_c
=8\pi e_q^2\alpha_sC\left[\frac{s(p^2-Q^2)}{tu}
\right]\;.
\end{eqnarray}
The complete expression is then written as 
\begin{eqnarray}
-g^{\mu\nu}{\hat W}_{\mu\nu}
&=&(-g^{\mu\nu}{\hat W}_{\mu\nu})_a+(-g^{\mu\nu}{\hat W}_{\mu\nu})_b+
(-g^{\mu\nu}{\hat W}_{\mu\nu})_c\;,
\nonumber \\
&=&8\pi e_q^2\alpha_sC\left[\frac{u}{t}+\frac{t}{u}+
\frac{p^2Q^2}{t^2}+\frac{p^2Q^2}{u^2}+\frac{2s(p^2-Q^2)}{tu}\right]\;.
\label{tm}
\end{eqnarray}

Next we calculate the contraction $P^\mu P^\nu W_{\mu\nu}$.
Note that the proton momentum $P$ is not parallel to the gluon momentum
$p$, because the gluon is off shell, while the proton is on shell.
In the center-of-mass frame $P$ possesses different components from those 
in the infinite-momentum frame. From the invariants,
\begin{eqnarray}
& &P\cdot p=P_0p_0-P_3|{\bf p}|=0\;,
\nonumber \\
& &x=z\xi=\frac{Q^2}{2P\cdot q}=\frac{Q^2}{2(P_0q_0+P_3|{\bf q}|)}\;,
\end{eqnarray}
the components $P_0$ and $P_3$ in the center-of-mass frame are solved to
be
\begin{eqnarray}
P_0&=&\frac{Q}{2\xi}\frac{1}{\sqrt{z(1-z+z p^2/Q^2)}}\;,
\nonumber \\
P_3&=&\frac{Q}{2\xi}\frac{1+2zp^2/Q^2}{\sqrt{z(1-z+z p^2/Q^2)}
\sqrt{1+4z^2p^2/Q^2}}\;.
\end{eqnarray}
Because of $P^2=0$, it is easy to obtain the transverse components
\begin{equation}
P_T=\frac{p_T^2}{\xi^2}\frac{1}{1+4z^2p^2/Q^2}\;.
\end{equation}

It will be found that all the contributions from Fig.~2 to 
$P^\mu P^\nu W_{\mu\nu}$ can be expressed in terms of the invariant,
\begin{equation}
P\cdot k=P_0k_0-P_3k_0\cos\theta-P_Tk_0\sin\theta\cos\phi\;,
\end{equation}
with
\begin{eqnarray}
P_0k_0&=&\frac{Q^2}{4x}\;,
\nonumber \\
P_3k_0&=&\frac{Q^2}{4x}\frac{1+2zp^2/Q^2}{\sqrt{1+4z^2p^2/Q^2}}\;,
\nonumber \\
P_Tk_0&=&\frac{p_TQ}{2x}\frac{\sqrt{z(1-z+z p^2/Q^2)}}
{\sqrt{1+4z^2p^2/Q^2}}\;,
\end{eqnarray}
$\phi$ being the azimuthal angle of the momentum ${\bf k}$.
Following a similar procedure, we derive the squared amplitudes
from Figs.~2(a), 2(b) and 2(c),
\begin{eqnarray}
\left(P^\mu P^\nu {\hat W}_{\mu\nu}\right)_a
&=&8\pi e_q^2\alpha_sC\frac{p^2}{t^2}\left[2(P\cdot k)^2
-\frac{Q^2}{z}P\cdot k\right]\;,
\nonumber \\
\left(P^\mu P^\nu {\hat W}_{\mu\nu}\right)_b
&=&\left(P^\mu P^\nu {\hat W}_{\mu\nu}\right)_a(t\leftrightarrow u)\;,
\nonumber \\
\left(P^\mu P^\nu W_{\mu\nu}\right)_c
&=&-8\pi e_q^2\alpha_sC\frac{1}{tu}
\biggl\{2(Q^2+p^2)(P\cdot k)^2
\nonumber \\
& &\hspace{2.0cm}
-\frac{Q^2}{z}\left[Q^2+p^2-\frac{Q^2}{2z}A\cos\theta\right]P\cdot k
\nonumber \\
& &\hspace{2.0cm}
\frac{Q^4}{4z^2}\left[\frac{Q^2}{2z}(1-A\cos\theta)+p^2\right]\biggr\}\;,
\label{tm2}
\end{eqnarray}
respectively, with $A=\sqrt{1+4z^2p^2/Q^2}$. The complete expression is
given by 
\begin{eqnarray}
P^\mu P^\nu {\hat W}_{\mu\nu}=
\left(P^\mu P^\nu {\hat W}_{\mu\nu}\right)_a+
\left(P^\mu P^\nu {\hat W}_{\mu\nu}\right)_b+
\left(P^\mu P^\nu {\hat W}_{\mu\nu}\right)_c\;.
\end{eqnarray}

To extract the hard scattering subamplitudes, we integrate
Eqs.~(\ref{tm}) and (\ref{tm2}) over phase space. For
$-g^{\mu\nu}{\hat W}_{\mu\nu}$ which is independnet of the azimuthal
angle $\phi$, the integral is written as
\begin{eqnarray}
H_2&=&\frac{1}{2\pi}\frac{1}{(2\pi)^4}\int d^4k' d^4k 
\delta^4(p+q-k-k') 
\nonumber \\
& &\times2\pi\delta_+(k^2) 2\pi\delta_+(k'^2)
z(-g^{\mu\nu}{\hat W}_{\mu\nu})\;,
\nonumber\\
&=&\frac{1}{32\pi^2}\int d\cos\theta 
z(-g^{\mu\nu}{\hat W}_{\mu\nu})\;.
\label{hg}
\end{eqnarray}
Substituting Eq.~(\ref{tm}) into Eq.~(\ref{hg}) and integrating it over 
$\cos\theta$, we obtain Eq.~(\ref{hgc}), where the relation $p^2=-p_T^2$ 
has been inserted. For another contraction $P^\mu P^\nu W_{\mu\nu}$, which
depends on $\phi$, the integral is given by 
\begin{equation}
H_L=\frac{1}{64\pi^3}\int d\cos\theta d\phi 
2z\frac{4x^2}{Q^2}P^\mu P^\nu {\hat W}_{\mu\nu}\;.
\label{hpp}
\end{equation}
After a tedious calculation, we obtain Eqs.~(\ref{ppab}) and (\ref{ppc}).

\newpage
\centerline{\large \bf Figure Captions}
\vskip 0.5cm
\noindent
{\bf Fig. 1.} (a) The derivative $-xdF/dx$ in axial gauge.
(b) The soft structure and (c) the ultraviolet structure of the
$O(\alpha_s)$ subdiagram containing the special vertex.
\vskip 0.5cm

\noindent
{\bf Fig. 2.} The quark box diagrams contributing to the hard scattering
subamplitudes $H_{2,L}$.
\vskip 0.5cm

\noindent
{\bf Fig. 3.} The dependence of $F_2$ on $x$ for $Q^2=8.5$, 12, 15, 25,
and 50 GeV$^2$ derived from the modified BFKL equation, the conventional
BFKL equation, and the CCFM equation.
The HERA data \cite{H1} are also shown.
\vskip 0.5cm

\noindent
{\bf Fig. 4.} The dependence of $F_L$ on $x$ for $Q^2=8.5$, 12, 15, 25,
and 50 GeV$^2$ derived from the modified BFKL equation, the conventional
BFKL equation, and the CCFM equation.
The HERA data \cite{H12} are also shown.
\vskip 0.5cm

\noindent
{\bf Fig. 5.} The dependence of $xg$ on $x$ derived from the modified BFKL 
equation.

\end{document}